\documentclass[11pt]{article}
\usepackage{amsmath,amssymb,color,epsfig,cite}
\usepackage{graphicx}
\usepackage{setspace}
\usepackage{mathrsfs}

\usepackage[english]{babel}
\usepackage{inputenc}
\usepackage{amsfonts,amsmath,amssymb}
\usepackage{mathrsfs}

\usepackage{graphicx}
\usepackage{color}
\usepackage{soul} 
\usepackage{hyperref}

\usepackage{subfigure}
\usepackage[normalem]{ulem}
\usepackage{array}
\usepackage{authblk}

\usepackage{amsfonts}
\newcommand{\hoch}[1]{$\, ^{#1}$}

\makeatletter
\@addtoreset{equation}{section}
\makeatother
\renewcommand{\theequation}{\thesection.\arabic{equation}}

\newcommand{\be}{\begin{equation}}
\newcommand{\ee}{\end{equation}}
\newcommand{\bea}{\setlength\arraycolsep{2pt} \begin{eqnarray}}
\newcommand{\eea}{\end{eqnarray}}

\def\0{{\sst{(0)}}}
\def\1{{\sst{(1)}}}
\def\2{{\sst{(2)}}}
\def\3{{\sst{(3)}}}
\def\4{{\sst{(4)}}}
\def\5{{\sst{(5)}}}
\def\6{{\sst{(6)}}}
\def\7{{\sst{(7)}}}
\def\8{{\sst{(8)}}}
\def\sst#1{{\scriptscriptstyle #1}}

\begin{document}

\vspace{15pt}
{\Large {\bf DeWitt boundary condition in one-loop quantum cosmology} }

\vspace{15pt}
{\bf Giampiero Esposito\hoch{1,2}}

\vspace{10pt}

\hoch{1}{\it Dipartimento di Fisica ``Ettore Pancini'', \\
Complesso Universitario di Monte S. Angelo, Via Cintia Edificio 6, 80126 Napoli, Italy.}

\hoch{2} {\it INFN Sezione di Napoli, \\
Complesso Universitario di Monte S. Angelo, Via Cintia Edificio 6, 80126 Napoli, Italy.}

\vspace{20pt}

\underline{ABSTRACT}

DeWitt's suggestion that the wave function of the universe 
should vanish at the classical big-bang singularity
is here considered within the framework of one-loop quantum 
cosmology. For pure gravity at one loop about a flat four-dimensional
background bounded by a 3-sphere, three choices of boundary 
conditions are considered: vanishing of the linearized magnetic
curvature when only transverse-traceless gravitational modes are quantized;
a one-parameter family of mixed boundary conditions for gravitational
and ghost modes; diffeomorphism invariant boundary conditions for
metric perturbations and ghost modes. A positive $\zeta(0)$ value
in these cases ensures that, when the 3-sphere boundary approaches
zero, the resulting one-loop wave function approaches zero. This
property may be interpreted by saying that, in the limit of small
three-geometry, the resulting one-loop wave function describes a
singularity-free universe. This property holds for one-loop
functional integrals, which are not necessarily equivalent
to solutions of the quantum constraint equations.

\noindent

\thispagestyle{empty}

\vfill
gesposito@na.infn.it.

\pagebreak

\section{Introduction}
\setcounter{equation}{0}

After the birth of relativistic cosmology thanks to Friedmann's
work \cite{F00}, and the subsequent
proof of the singularity theorems of Penrose, Hawking 
and Geroch \cite{F01,F02,F03,F04,F05,F06}, 
it became well accepted by the scientific community
that classical general relativity leads to the occurrence of
cosmological singularities (a spacetime being singular if it is timelike or
null geodesically incomplete) in a generic way. Since then, several
developments occurred, and in particular we here mention what follows.
\vskip 0.3cm
\noindent
(i) At classical level, the work of Christodoulou and Klainerman \cite{F07}
led to the discovery of asymptotically flat spacetimes which are
timelike geodesically complete.
\vskip 0.3cm
\noindent
(ii) At quantum level, DeWitt \cite{F08} proposed to look at the behaviour
of the wave function of the universe in correspondence with the
classical big-bang singularity. Such a proposal was assessed 
in the outstanding work of Ref. \cite{F09}
\vskip 0.3cm
\noindent
(iii) Along many years, various concepts of singularity were
conceived, as can be seen in an important review of 
Kamenshchik \cite{F10}.

Moreover, in the literature on quantum gravity and quantum 
cosmology, several approaches were developed to study the possible
quantum origin of spacetime geometry. One-loop effects in the
early universe were investigated in detail, especially with the
help of $\zeta$-function methods. It is the aim of our review to
describe them and then discuss their relevance for the singularity
issue in cosmology. The structure of the paper is as follows.

Section $2$ presents in detail a $\zeta(0)$ calculation when only
transverse-traceless perturbations are considered, with boundary
conditions requiring the vanishing of linearized magnetic curvature
on the 3-sphere boundary. Section $3$ discusses a one-parameter 
family of $\zeta(0)$ values obtained with mixed boundary
conditions for metric perturbations and ghost fields. Sections
$4-7$ outline the basic steps of the $\zeta(0)$ calculation with 
diffeomorphism invariant boundary conditions. Open problems are
discussed in Sec. $8$.

\section{Linearized magnetic curvature vanishing on $S^3$}
\setcounter{equation}{0}

We study pure gravity at one loop about flat Euclidean 4-space
with a 3-sphere boundary of radius $a$, because when 
$a \rightarrow 0$ this is the limiting case of a 4-sphere
geometry bounded by a 3-sphere \cite{F11}. 
The prefactor of the semiclassical
wave function is given by ($I_{2}$ denoting the part of the
action quadratic in metric perturbations)
\begin{equation}
P(a)=\int e^{-I_{2}[\gamma]} \; D\gamma,
\label{(2.1)}
\end{equation}
which is a functional integral over all metric perturbations 
$\gamma_{ab}$ that are regular at the origin $\tau=0$ and satisfy
a given boundary condition at $\tau=a$. Integration is here
restricted to the physical degrees of freedom, which are found
by using the Hamiltonian formulation with the following 
transverse-traceless choice of gauge condition: 
\begin{equation}
\sum_{i}(D^{i}\gamma)_{ij}=0, \;
\sum_{k}\gamma_{k}^{k}=0.
\label{(2.2)}
\end{equation}
These relations pick out the transverse-traceless tensor
hyperspherical harmonics $G_{ij}^{(n)}(\phi^{k})$ multiplied
by functions of the radial coordinate $\tau$. Hence we write
\begin{equation}
\gamma_{ij}=\gamma_{ij}^{TT}=\sum_{n=3}^{\infty}q^{n}(\tau)
G_{ij}^{(n)}(\phi^{k}).
\label{(2.3)}
\end{equation}
Our work in Ref. \cite{F12} studied the Breitenlohner-Freedman-Hawking
\cite{F13,F14}
local boundary conditions for fields of spin 
$0,{1\over 2},1,{3\over 2},2$. For gravity, these imply that
the linearized magnetic curvature should vanish at the
boundary. Our detailed analysis in Sec. 7.3 of Ref. \cite{F12}
never appeared in journals, and hence it is of interest for
our review article.

The action which is quadratic in metric perturbations involves
a second-order elliptic operator $A$ 
with eigenvalues $\lambda_{n}$
for which one can define a spectral $\zeta$-function
\begin{equation}
\zeta_{A}(s)={\rm Tr}_{L^2} A^{-s}
=\sum_{n}(\lambda_{n})^{-s}.
\label{(2.4)}
\end{equation}
Eventually, as was shown by Schleich \cite{F11}, the 
prefactor of the 
semiclassical wave function of Eq. (2.1), with $\gamma$
having the form (2.3), can be expressed as 
\begin{equation}
P(a)={1 \over \sqrt{{\rm det}\left(
{-\nabla_{f}\nabla^{f}\over 4\pi l_{p}^{2}\mu^{2}}
\right)}}
=D \; a^{\zeta(0)},
\label{(2.5)}
\end{equation}
where $-\nabla_{f}\nabla^{f}$ is the Laplacian acting
on transverse-traceless metric perturbations, and
$\zeta(0)$ is the value at $s=0$ of the analytic
continuation of the spectral $\zeta$-function (2.4) 
(for further details, see now the Appendix on the 
one-loop approximation).
Thus, within a functional-integral framework, the wave 
function of the univere may fulfill the DeWitt boundary
condition if and only if $\zeta(0)$ is positive.

The linearized magnetic curvature for gravity 
is defined from the Weyl tensor $C$ and from the normal $n$
to the boundary according to the rule (with summation over
repeated tensor indices)
$$
B_{ij} \equiv {1 \over 2}\varepsilon_{j \mu}^{\; \; \; kl}
C_{kli \nu}n^{\mu}n^{\nu},
$$
and it can only vanish on $S^{3}$ if \cite{F12}                   
\begin{equation}
\sum_{n=3}^{\infty}{dq^{n}\over d\tau}(a)\epsilon_{j}^{\; \; kl}
\left(G_{il\mid k}^{(n)} -G_{ik\mid l}^{(n)}\right)=0  .
\label{(2.6)}
\end{equation}
The only condition on the modes
which ensures the validity of (2.6) is 
\begin{equation}
{dq^{n}\over d\tau}(a)=0  ,  \; \forall n \geq 3 .
\label{(2.7)}
\end{equation}
We are now interested in evaluating $\zeta(0)$ by using (2.7). Thus,
after setting $\tau=t$, we study 
the kernel of the heat equation for the operator 
\begin{equation}
P_{n} \equiv -\left({d^{2}\over dt^{2}}-{1\over t}{d\over dt}
-{(n^{2}-1)\over t^{2}}\right)
 ,  \; \forall n \geq 3  ,
\label{(2.8)}
\end{equation}
which results from studying the Laplacian on transverse-traceless
metric perturbations.
On denoting by $E>0$ the eigenvalues of $P_{n}$, we find that
its eigenfunctions regular at the origin are (up to a
multiplicative constant) 
\begin{equation}
u_{n}(t)=t J_{n}(\sqrt{E}t)=q^{n}(t)  .
\label{(2.9)}
\end{equation}
Thus, the boundary condition (2.7) implies the eigenvalue condition 
\begin{equation}
J_{n}(\sqrt{E}a)+\sqrt{E}a {\dot J}_{n}(\sqrt{E}a)=0  ,        
\; \forall n \geq 3  .
\label{(2.10)}
\end{equation}
This equation is of the general kind studied in Ref. 
\cite{F15}. Setting now $a=1$ for simplicity, we define the function 
\begin{equation}
F_{n}(z)\equiv J_{n}(z)+z{\dot J}_{n}(z)  , \;     
\forall n \geq 3  .
\label{(2.11)}
\end{equation}
Of course, the consideration of such $F_{n}(z)$ is suggested by (2.10).
It only has real simple zeros apart from $z=0$ (page 482 of Ref. \cite{F16}).
The basic idea is now the following \cite{F15}. Given the 
zeta-function at large $x$ 
\begin{equation}
\zeta(s,x^{2}) \equiv 
\sum_{n}{\Bigr(\lambda_{n}+x^{2}\Bigr)}^{-{\textstyle s}},
\label{(2.12)}
\end{equation}
one has in four dimensions (see theorem 2 on page 6 of Ref. \cite{F17}) 
\begin{equation}
\Gamma(3)\zeta(3,x^{2})=\int_{0}^{\infty}t^{2}e^{-x^{2}t}G(t)\; dt
\sim \sum_{n=0}^{\infty}B_{n}\Gamma\left(1+{n\over 2}\right)x^{-n-2},
\label{(2.13)}
\end{equation}
where we have used the asymptotic expansion 
of the heat kernel for $t \rightarrow 0^{+}$, i.e.
\begin{equation}
G(t)\sim \sum_{n=0}^{\infty}B_{n}t^{{n\over 2}-2}.
\label{(2.14)}
\end{equation}
Strictly speaking, since we have not proved general results on the
existence of the asymptotic expansion of the heat kernel, our formula
(2.14) could be initially regarded as an assumption. However, existence
theorems hold for the problems studied in this paper 
(\cite{F18,F19}).

On the other hand, one also has the identity :
\begin{equation}
\Gamma(3)\zeta(3,x^{2})=-\sum_{p=0}^{\infty}N_{p}{\left(-{1\over 2x}
{d\over dx}\right)}^{3}\log \Bigr((ix)^{-p}F_{p}(ix)\Bigr),
\label{(2.15)}
\end{equation}
where $N_{p}$ is the degeneracy of the problem. Thus the comparison of
(2.13) and (2.15) can yield the coefficients $B_{n}$ and in particular
$\zeta(0)=B_{4}$, provided we carefully perform a uniform Debye expansion
of $F_{p}(ix)$. It should be emphasized that this technique seems to be the
most efficient. In fact, by using this algorithm
Moss \cite{F15} has been able to compute $\zeta(0)$
for a real scalar field subject to Robin boundary conditions, whereas the
technique of Kennedy based on charge layers on the plane tangent to $S^3$
failed to provide such a value \cite{F20,F21}. Indeed,
the eigenvalue condition (2.10) is of the Robin type (just set $\beta=1$
in Eq. (22) of Ref. \cite{F15}). Thus, on passing to the variable
$x \rightarrow ix$ and then defining
$\alpha_{p} \equiv \sqrt{p^{2}+x^{2}}$,
$C \equiv -\log(\sqrt{2\pi})$, we can write 
\begin{equation}
\log \Bigr((ix)^{-p}F_{p}(ix)\Bigr) \sim C 
-p \log(p+\alpha_{p})
+{1\over 2}\log(\alpha_{p})
+ \alpha_{p} + \sum_{n=1}^{\infty}\sum_{r=0}^{n}a_{nr}p^{2r}
\alpha_{p}^{-n-2r}  .
\label{(2.16)}
\end{equation}
The coefficients $a_{nr}$ in (2.16) can be computed by comparison using
the formula 
\begin{equation}
\sum_{n=1}^{\infty}\sum_{r=0}^{n}a_{nr}t^{2r}=\sum_{m=1}^{\infty}a_{m}(t),
\label{(2.17)}
\end{equation}
because the $a_{m}(t)$ are known polynomials in $t$ arising from uniform 
asymptotic expansions of Bessel functions and their first derivatives.
Thus, setting $\beta=1$ in the formulae (29-31)
of Ref. \cite{F15} for the $a_{m}(t)$, we find in our case that 
\begin{equation}
a_{10}={5\over 8}  , \;  a_{11}={7\over 24}  ,
\label{(2.18)}
\end{equation}
\begin{equation}
a_{20}=-{3\over 16}  , \; 
a_{21}={1\over 8}  , \; 
a_{22}=-{7\over 16} ,
\label{(2.19)}
\end{equation}
\begin{equation}
a_{30}={17\over 384} , \;  a_{31}={389\over 640}, \;
a_{32}=-{203\over 128} , \;  a_{33}={1463\over 1152},
\label{(2.20)}
\end{equation}
plus infinitely many other coefficients that we do not strictly need here. We
can now insert (2.16)-(2.20) into (2.15), apply three times the differential
operator $-{1\over 2x}{d\over dx}$, and finally use the contour formula
for positive integer values of $k$ \cite{F15} 
\begin{equation}
\sum_{p=0}^{\infty}p^{2k}\alpha_{p}^{-2k-m}=
{{\Gamma \left(k+{1\over 2}\right)\Gamma \left({m\over 2}-{1\over 2}\right)}
\over 2\Gamma \left(k+{m\over 2}\right)}x^{1-m}
 , \;  \forall k=1,2,3,...  \; ,
\label{(2.21)}
\end{equation}
and the known properties of the $\Gamma$-function \cite{F22}.
Now, writing the asymptotic expansion of the right-hand side of (2.15)
in the form 
\begin{equation}
\Gamma(3)\zeta(3,x^{2}) \sim \sum_{n=0}^{\infty}b_{n}x^{-n-2},
\label{(2.22)}
\end{equation}
the comparison with (2.15) shows that 
\begin{equation}
\zeta(0)=B_{4}={b_{4}\over 2}=\zeta^{I}(0)+\zeta^{II}(0)  ,
\label{(2.23)}
\end{equation}
since it is well-known that the asymptotic expansion, if it exists, is unique.
The two contributions to $\zeta(0)$ are obtained from the following formulae:
\begin{equation}
\Gamma(3)\zeta(3,x^{2}) \sim
\Bigr[\sigma_{1}+\sigma_{2}\Bigr] \sim
\sum_{n=0}^{\infty}b_{n}x^{-n-2},                       
\label{(2.24)}
\end{equation}
\begin{equation}
\sigma_{1} \sim
-\sum_{p=0}^{\infty}N_{p}{\left(-{1\over 2x}{d\over dx}\right)}^{3}
\left[-p \log(p+\alpha_{p})
+{1\over 2}\log(\alpha_{p})+\alpha_{p}\right]  ,
\label{(2.25)}
\end{equation}
\begin{equation}
\sigma_{2} \sim
-\sum_{p=0}^{\infty}N_{p}{\left(-{1\over 2x}{d\over dx}\right)}^{3}
\sum_{n=1}^{\infty}\sum_{r=0}^{n}a_{nr}p^{2r}\alpha_{p}^{-n-2r}  .
\label{(2.26)}
\end{equation}
Bearing in mind (2.15)-(2.16), we write (2.24)-(2.26) because we can apply
theorem 3 on page 7 of Ref. \cite{F17}, concerning the differentiation of
asymptotic expansions.

Thus $\zeta^{I}(0)$ (respectively $\zeta^{II}(0)$) is half     
the coefficient of $x^{-6}$ in the asymptotic expansion
of $\sigma_{1}$ (respectively $\sigma_{2}$). We first study the asymptotic
expansion of $\sigma_{2}$, since it is easier to perform this calculation.
In our problem the degeneracy $N_{p}$ is \cite{F11} 
\begin{equation}
N_{p}=0 \;  \forall p=0,1,2  , \;  N_{p}=2(p^{2}-4)  \;
\forall p\geq 3 .
\label{(2.27)}
\end{equation}
This is why we find 
\begin{equation}
\sigma_{2} \sim
-\sum_{n=1}^{\infty}\sum_{r=0}^{n}a_{nr}\left(r+{n\over 2}\right)
\left(r+{n\over 2}+1\right)\left(r+{n\over 2}+2\right)               
\Bigr[(G-H)(r,x,n)\Bigr]  ,
\label{(2.28)}
\end{equation}
where, setting $A=-8,\; B=2$ (cf. (2.27)) we have, using also (2.21),
\begin{eqnarray}
G(r,x,n)&=& \sum_{p=0}^{\infty}(A+Bp^{2})p^{2r}\alpha_{p}^{-n-2r-6}   
= {\rm O}(x^{-n-6})
\nonumber \\
&+& {A\over 2}{\Gamma \left(r+{1\over 2}\right)
\Gamma \left({n\over 2}+{5\over 2}\right)\over
\Gamma \left(r+{n\over 2}\right)}
{x^{-5-n}\over {\left(r+{n\over 2}\right)\left(r+{n\over 2}+1\right)
\left(r+{n\over 2}+2\right)}}
\nonumber \\
&+& {B\over 2}{\Gamma \left(r+{3\over 2}\right)
\Gamma \left({n\over 2}+{3\over 2}\right)\over
\Gamma \left(r+{n\over 2}\right)}
{x^{-3-n}\over {\left(r+{n\over 2}\right)\left(r+{n\over 2}+1\right)
\left(r+{n\over 2}+2\right)}}  ,
\label{(2.29)}
\end{eqnarray}
\begin{eqnarray}
H(r,x,n)&=& \sum_{p=0}^{2}2(p^{2}-4)p^{2r}\alpha_{p}^{-n-2r-6}   
=-6x^{-n-2r-6}{\left(1+{1\over x^{2}}\right)}^{-{n\over 2}-r-3}
\nonumber \\
&-& 8 \delta_{r0}x^{-n-6}  .
\label{(2.30)}
\end{eqnarray}
Thus $H(r,x,n)$ gives rise to terms in (2.28) which contain
$x^{-k}$ with $k\geq 7$, and does not contribute to $\zeta^{II}(0)$. This is
why (2.28)-(2.29) lead to 
\begin{equation}
\zeta^{II}(0)={1\over 2}\Bigr[-A(a_{10}+a_{11})-B(a_{30}+a_{31}+a_{32}
+a_{33})\Bigr]  .
\label{(2.31)}
\end{equation}
The insertion of (2.18), (2.20) and (2.27) into (2.31) finally
yields 
\begin{equation}
\zeta^{II}(0)={11\over 3}-{121\over 360}={1199\over 360}.
\label{(2.32)}
\end{equation}

The calculation of (2.25) is more involved. By performing the three
derivatives and using the identity ${1\over 2x}{d\alpha_{p} \over dx}=      
{1\over 2\alpha_{p}}$, we find 
\begin{equation}
{\left({1\over 2x}{d\over dx}\right)}^{3}
\log \left({1\over {p+\alpha_{p}}}\right)
=(p+\alpha_{p})^{-3}\left[-\alpha_{p}^{-3}
-{9\over 8}p\alpha_{p}^{-4}-{3\over 8}
p^{2}\alpha_{p}^{-5}\right]  .
\label{(2.33)}
\end{equation}
This intermediate step is very important because it proves that by summing
over all integer values of
$p$ from $0$ to $\infty$ we get a convergent series. However, to be  
able to perform the $\zeta(0)$ calculation, it is convenient                  
to use the identity 
\begin{equation}
(p+\alpha_{p})^{-3}={(\alpha_{p} -p)^{3}\over x^{6}}  .
\label{(2.34)}
\end{equation}
Upon inserting (2.34) into (2.33) and re-expressing $p^{2}$ as
$\alpha_{p}^{2}-x^{2}$, we obtain 
\begin{eqnarray}
\; & \; & 
{\left({1\over 2x}{d\over dx}\right)}^{3}
\Bigr[-p\log(p+\alpha_{p})\Bigr]=
-px^{-6}+p^{2}x^{-6}\alpha_{p}^{-1}
+{p^{2}\over 2}x^{-4}\alpha_{p}^{-3}+{3\over 8}
p^{2}x^{-2}\alpha_{p}^{-5}
\nonumber \\
& \equiv & M(x,\alpha_{p},p)  ,
\label{(2.35)}
\end{eqnarray}
which implies 
\begin{equation}
\sigma_{1} \sim \left[\sum_{p=0}^{\infty}N_{p}M(x,\alpha_{p},p)\right]
+\sigma''_{1} \sim \Bigr[\sigma'_{1}+\sigma''_{1}\Bigr]  ,
\label{(2.36)}
\end{equation}
where 
\begin{eqnarray}
\sigma''_{1}&=& -\sum_{p=0}^{\infty}N_{p}\left(-{\alpha_{p}^{-6}\over 2}-
{3\over 8}\alpha_{p}^{-5}\right) \nonumber \\
&=& \sum_{p=0}^{\infty}(A+Bp^{2})
\left({\alpha_{p}^{-6}\over 2}+
{3\over 8}\alpha_{p}^{-5}\right) \nonumber \\
&+& \sum_{p=0}^{2}(A+Bp^{2})\left(-{\alpha_{p}^{-6}\over 2}-{3\over 8}
\alpha_{p}^{-5}\right)  .
\label{(2.37)}
\end{eqnarray}
The infinite sum on the right-hand side of (2.37) contributes to $\zeta(0)$
only through the following part:
\begin{equation}
\sum_{p=0}^{\infty}{A\over 2}\alpha_{p}^{-6} = {A\over 2}
\left[{x^{-6}\over 2}+{\pi \over 2}{3!! \over 4!!}x^{-5}\right].
\label{(2.38)}
\end{equation}
The result (2.38) is proved by applying the Euler-Maclaurin formula 
\cite{F17} to the calculation of $\sum_{p=0}^{\infty}
(p^{2}+x^{2})^{-3}$, and then using the formula (3.249.1) on page 294
of Ref. \cite{F23}.
Also the finite sum on the right-hand side of (2.37) contributes
to $\zeta(0)$. In fact one finds (we have $x\rightarrow \infty$) :
\begin{eqnarray}
\; & \; & 
\sum_{p=0}^{2}(A+Bp^{2})\left(-{\alpha_{p}^{-6}\over 2}
-{3\over 8}\alpha_{p}^{-5}
\right)=
-\left({A\over 2}+{B\over 2}\right)x^{-6}\left[1-{3\over x^{2}}
+{6\over x^{4}}+...\right] \nonumber \\
&-& {A\over 2}x^{-6}-{3\over 8}Ax^{-5} \nonumber \\
&-& {3\over 8}(A+B)x^{-5}\left[1-{5\over 2x^{2}}+{35\over 8x^{4}}+...\right],
\label{(2.39)}
\end{eqnarray}
which implies that the total contribution of $\sigma''_{1}$ to $\zeta(0)$
is given by 
\begin{equation}
\zeta^{Ib}(0)={1\over 2}\left(-A-{B\over 2}\right)+{A\over 8}=
{7\over 2}-1={5\over 2}  .
\label{(2.40)}
\end{equation}
Thus, we have so far 
\begin{equation}
\zeta(0)=\zeta^{I}(0)+\zeta^{II}(0)=\zeta^{Ia}(0)+\zeta^{Ib}(0)+\zeta^{II}(0),
\label{(2.41)}
\end{equation}
where 
\begin{equation}
\zeta^{Ib}(0)+\zeta^{II}(0)={5\over 2}+{1199\over 360}  .
\label{(2.42)}
\end{equation}
It now remains to compute $\zeta^{Ia}(0)$, i.e. the contribution to 
$\zeta(0)$ due to $\sigma'_{1}$ in (2.36). Indeed, one has 
\begin{equation}
\sigma'_{1} \sim \left[
A\sum_{p=0}^{\infty}M(x,\alpha_{p},p)+B\sum_{p=0}^{\infty}
p^{2}M(x,\alpha_{p},p)
-\sum_{p=0}^{2}(A+Bp^{2})M(x,\alpha_{p},p)\right]  .
\label{(2.43)}
\end{equation}
Let us now denote by $\Sigma^{(a)}$, $\Sigma^{(b)}$ and
$\Sigma^{(c)}$ the three sums on the right-hand side of (2.43).
Both $\Sigma^{(a)}$ and $\Sigma^{(b)}$ contain divergent parts in view
of (2.35). These {\it fictitious}
divergences may be regularized dividing by $\alpha_{p}^{2s}$ and
then taking the limit as $s$ tends to zero, as shown in Ref. \cite{F15}.
It might not appear {\it a priori} obvious that this technique leads to
unambiguous results, since the limit as $s \rightarrow 0$ is a delicate
mathematical point. However, a fundamental consistency check is presented
in Sec. 7.4 of Ref. \cite{F12} for all one-loop calculations involving only 
physical degrees of freedom of bosonic fields,
showing that the method is correct. In performing
the calculation we must use the contour formula (2.21) and also the
following asymptotic expansion \cite{F15}:
\begin{equation}
\sum_{p=0}^{\infty}p\alpha_{p}^{-1-n}\sim {x^{1-n}\over \sqrt{\pi}}
\sum_{r=0}^{\infty}{2^{r}\over r!}{\widetilde B}_{r}x^{-r}
{{\Gamma\left({r\over 2}+{1\over 2}\right)
\Gamma\left({n\over 2}-{1\over 2}+{r\over 2}\right)}\over
{2\Gamma\left({1\over 2}+{n\over 2}\right)}}\cos\left({r\pi\over 2}\right),
\label{(2.44)}
\end{equation}
where ${\widetilde B}_{0}=1$, ${\widetilde B}_{1}=-{1\over 2}$,
${\widetilde B}_{2}={1\over 6}$,
${\widetilde B}_{4}=-{1\over 30}$ etc. are Bernoulli numbers. Thus,
using the label $R$ for the regularized quantities, we define 
\begin{eqnarray}
\Sigma_{R}^{(a)} &\equiv & A\Biggr[-x^{-6}\left(\lim_{s \to 0}\sum_{p=0}^{\infty}
p\alpha_{p}^{-1-(2s-1)}\right)   
+x^{-6}\left(\lim_{s \to 0}\sum_{p=0}^{\infty}
p^{2}\alpha_{p}^{-2-(2s-1)}\right) \nonumber \\
&+& {x^{-4}\over 2}\left(\lim_{s \to 0}\sum_{p=0}^{\infty}
p^{2}\alpha_{p}^{-2-(2s+1)}\right)   
+{3\over 8}x^{-2}\left(\lim_{s \to 0}\sum_{p=0}^{\infty}p^{2}
\alpha_{p}^{-2-(2s+3)}\right)\Biggr].
\label{(2.45)}
\end{eqnarray}
In view of (2.44), the first limit in (2.45) gives the following
contribution to $\zeta(0)$:
\begin{equation}
\delta_{1}=-{A\over 2}\left(-{{\widetilde B}_{2}\over \sqrt{\pi}}
\Gamma \left({3\over 2}\right)\right)={A\over 24}=-{1\over 3},
\label{(2.46)}
\end{equation}
whereas the other limits in (2.45) do not contribute to $\zeta(0)$ in
view of (2.21), because one only gets terms proportional to $x^{-4}$.

Moreover, bearing in mind the identity 
\begin{equation}
\sum_{p=0}^{\infty}p^{3}\alpha_{p}^{-2s}=
\sum_{p=0}^{\infty}p\alpha_{p}^{-1-(2s-3)}
-x^{2}\sum_{p=0}^{\infty}p\alpha_{p}^{-1-(2s-1)} ,
\label{(2.47)}
\end{equation}
we also define 
\begin{eqnarray}
\Sigma_{R}^{(b)}&\equiv & B\Biggr[-x^{-6}\left(\lim_{s \to 0}\sum_{p=0}^{\infty}
p^{3}\alpha_{p}^{-2s}\right)
+x^{-6}\left(\lim_{s \to 0}\sum_{p=0}^{\infty}
p^{4}\alpha_{p}^{-4-(2s-3)}\right)
\nonumber \\
&+& {x^{-4}\over 2}\left(\lim_{s \to 0}\sum_{p=0}^{\infty}p^{4}
\alpha_{p}^{-4-(2s-1)}\right)
+{3\over 8}x^{-2}\left(\lim_{s \to 0}\sum_{p=0}^{\infty}
p^{4}\alpha_{p}^{-4-(2s+1)}\right)\Biggr].
\label{(2.48)}
\end{eqnarray}
In view of (2.44) and (2.47), the first limit in (2.48) gives the
following contribution to $\zeta(0)$:
\begin{equation}
\delta_{2}=-{B\over 2}\left(-{{\widetilde B}_{4}\over 4}\right)=-{B\over 240}
=-{1\over 120}  .
\label{(2.49)}
\end{equation}
Note that the second sum in (2.47) does not contribute to
$\delta_{2}$ because its only constant term contains 
${\Gamma(s+1)\over \Gamma(s)}$, which tends to $0$ as $s \rightarrow 0$.
The other limits in (2.48) do not contribute to $\zeta(0)$ in view of
(2.21), because they only yield terms proportional to $x^{-2}$.

Last, the sum $\Sigma^{(c)}$ in (2.43) has the following
asymptotic behaviour as $x\rightarrow \infty$:
\begin{equation}
\Sigma^{(c)}\sim \left[(3A+9B)x^{-6}+\sum_{k=0}^{\infty}
(AC_{k}+BD_{k})x^{-7-k}\right]  ,
\label{(2.50)}
\end{equation}
which yields the following contribution to $\zeta(0)$:
\begin{equation}
\delta_{3}={(3A+9B)\over 2}=-3  .
\label{(2.51)}
\end{equation}
To sum up, we find 
\begin{equation}
\zeta^{Ia}(0)=\delta_{1}+\delta_{2}+\delta_{3}
=-{1\over 3}-{1\over 120}-3, 
\label{(2.52)}
\end{equation}
so that the full $\zeta(0)$ for physical degrees of freedom 
is given by (cf. (2.41)-(2.42))
\begin{equation}
\zeta(0)=\zeta^{Ia}(0)+{5\over 2}+{1199\over 360}={112\over 45}.
\label{(2.53)}
\end{equation}

\section{First example of mixed boundary conditions on the whole set
of metric perturbations and ghost modes}
\setcounter{equation}{0}

The previous example is very instructive, but of course it would
be desirable to compute the effect of boundary conditions on
the whole set of metric perturbations and 
Feynman-DeWitt-Faddeev-Popov ghost fields \cite{F24,F25,F26}. 
For this purpose, the 
work in Ref. \cite{F27} studied the following one-parameter family
of mixed boundary conditions ($\lambda$ being a freely
specifiable real parameter):
\begin{equation}
\left[{\partial h_{ij}\over \partial \tau}+{\lambda \over \tau}h{ij}
\right]_{\partial M}=0,
\label{(3.1)}
\end{equation}
\begin{equation}
\Bigr[h_{0i}\Bigr]_{\partial M}=0,
\label{(3.2)}
\end{equation}
\begin{equation}
\Bigr[h_{00}\Bigr]_{\partial M}=0,
\label{(3.3)}
\end{equation}
\begin{equation}
\left[{\partial \varphi_{i}\over \partial \tau}
+{\lambda \over \tau}\varphi_{i}\right]_{\partial M}=0,
\label{(3.4)}
\end{equation}
\begin{equation}
\left[{\partial \varphi_{0}\over \partial \tau}
+{(\lambda+1)\over \tau}\varphi_{0}\right]_{\partial M}=0.
\label{(3.5)}
\end{equation}
With our notation, $\tau$ lies in the closed interval 
$[0,a]$, $h_{ij},h_{0i},h_{00}$ are 
the components of metric perturbations,
$\varphi_{i}$ and $\varphi_{0}$ are covariant components
of the ghost field of quantum gravity.
One therefore deals with transverse-traceless modes, scalar modes,
vector modes, decoupled scalar modes, decoupled vector modes,
scalar ghost modes, vector ghost modes, decoupled ghost mode.

A one-parameter family of full $\zeta(0)$ values is therefore
obtained \cite{F27}:
\begin{equation}
\zeta_{\lambda}(0)={89 \over 90}+{\lambda \over 3}
(\lambda^{2}-9\lambda -3).
\label{(3.6)}
\end{equation}
The $\lambda$-dependent part of (3.6) is always positive either for all 
\begin{equation}
\lambda > {{9 + \sqrt{93}}\over 2},
\label{(3.7)}
\end{equation}
or for all 
\begin{equation}
\lambda \in \left ]{{9-\sqrt{93}}\over 2},0 \right [.
\label{(3.8)}
\end{equation}
Equations (3.7) and (3.8) are sufficient conditions for
positivity of the full $\zeta_{\lambda}(0)$, and other
suitable values of $\lambda$ can be computed numerically.

This model is more complete than the one in Sec. $2$,
since it deals with all perturbative modes in the one-loop
functional integral. However, it still suffers from a
non-trivial drawback: the whole set of boundary conditions
(3.1)-(3.5) is not completely invariant under infinitesimal
diffeomorphisms on metric perturbations. For this reason,
we resort to the boundary conditions of Sec. $4$.

\section{Completely diff-invariant boundary conditions}
\setcounter{equation}{0}

The boundary conditions that we study are part of a unified scheme for
Maxwell, Yang--Mills and Quantized General Relativity at one loop, 
i.e. \cite{F28,F29} 
\begin{equation}
\Bigr[\pi {\mathcal A}\Bigr]_{\mathcal B}=0,
\label{(4.1)}
\end{equation}
\begin{equation}
\Bigr[\Phi(A)\Bigr]_{\mathcal B}=0,
\label{(4.2)}
\end{equation}
\begin{equation}
[\varphi]_{\mathcal B}=0.
\label{(4.3)}
\end{equation}
With our notation, $\pi$ is a projector acting on the gauge field 
$\mathcal A$, $\Phi$ is the gauge-fixing functional, $\varphi$ is
the full set of ghost fields. Both
equation (4.1) and (4.2) are preserved under infinitesimal 
gauge transformations 
provided that the ghost obeys homogeneous Dirichlet conditions as 
in (4.3). For gravity, we choose $\Phi$ so as to have an operator $P$
of Laplace type on metric perturbations in the one-loop Euclidean theory.

\section{Eigenvalue conditions for scalar modes}
\setcounter{equation}{0}

On the Euclidean 4-ball, we expand metric perturbations $h_{\mu \nu}$
in terms of scalar, transverse vector, transverse-traceless tensor
harmonics on $S^{3}$. For vector, tensor and ghost modes, boundary
conditions reduce to Dirichlet or Robin. 
For scalar modes, one finds eventually the eigenvalues $E=x^{2}$ from
the roots $x$ of \cite{F30,F31} 
\begin{equation}
J_{n}'(x) \pm {n\over x}J_{n}(x)=0,
\label{(5.1)}
\end{equation}
\begin{equation}
J_{n}'(x)+\left(-{x\over 2}\pm {n\over x}\right)J_{n}(x)=0.
\label{(5.2)}
\end{equation}
Note that both $x$ and $-x$ solve the same equation. 

\section{Four spectral $\zeta$-functions for scalar modes}
\setcounter{equation}{0}

From Eqs. (5.1) and (5.2) we obtain the following integral representations
of the resulting $\zeta$-functions upon exploiting the Cauchy theorem and 
rotation of contour:
\begin{equation}
\zeta_{A,B}^{\pm}(s) \equiv {(\sin \pi s)\over \pi}
\sum_{n=3}^{\infty}n^{-(2s-2)}\int_{0}^{\infty}dz \; z^{-2s}
{\partial \over \partial z}\log F_{A,B}^{\pm}(zn),
\label{(6.1)}
\end{equation}
where (here $\beta_{+} \equiv n, \beta_{-} \equiv n+2$)
\begin{equation}
F_{A}^{\pm}(zn) \equiv z^{-\beta_{\pm}}\Bigr(znI_{n}'(zn)
\pm n I_{n}(zn)\Bigr),
\label{(6.2)}
\end{equation}
\begin{equation}
F_{B}^{\pm}(zn) \equiv z^{-\beta_{\pm}}\left(znI_{n}'(zn)
+\left({(zn)^{2}\over 2} \pm n \right)I_{n}(zn)\right),
\label{(6.3)}
\end{equation}
$I_{n}$ being the modified Bessel functions of first kind.
Regularity at the origin is easily proved in the elliptic sectors,
corresponding to $\zeta_{A}^{\pm}(s)$ and $\zeta_{B}^{-}(s)$.

\section{Regularity of $\zeta_{B}^{+}$ at $s=0$}
\setcounter{equation}{0}

We now define $T \equiv (1+z^{2})^{-1/2}$ and consider the uniform
asymptotic expansion (away from $T =1$)
\begin{equation}
z^{\beta_{+}}
F_{B}^{+}(zn) \sim {{\rm e}^{n\eta(T)}\over h(n)\sqrt{T}}
{(1-T^{2})\over T}
\left(1+\sum_{j=1}^{\infty}{r_{j,+}(T)\over n^{j}}\right),
\label{(7.1)}
\end{equation}
the functions $r_{j,+}$ being obtained from the Olver polynomials
for the uniform asymptotic expansion of $I_{n}$ and $I_{n}'$.
On splitting $\int_{0}^{1}dT=\int_{0}^{\mu}dT
+\int_{\mu}^{1}dT$ with $\mu$ small, we get an asymptotic expansion
of the l.h.s. by writing, {\it in the first interval} on the r.h.s.,  
\begin{equation}
\log \left(1+\sum_{j=1}^{\infty}{r_{j,+}(T)\over n^{j}}\right)
\sim \sum_{j=1}^{\infty}{R_{j,+}(T)\over n^{j}},
\label{(7.2)}
\end{equation}
and then computing
\begin{equation}
C_{j}(\tau) \equiv {\partial R_{j,+}\over \partial T}
=(1-T)^{-j-1}\sum_{a=j-1}^{4j}K_{a}^{(j)}T^{a}.
\label{(7.3)}
\end{equation}
The integral $\int_{\mu}^{1}dT$ is instead found to yield a vanishing
contribution in the $\mu \rightarrow 1$ limit.
Remarkably, by virtue of the spectral identity
\begin{equation}
g(j) \equiv \sum_{a=j}^{4j}{\Gamma(a+1)\over \Gamma(a-j+1)}
K_{a}^{(j)}=0,
\label{(7.4)}
\end{equation}
which holds $\forall j=1,...,\infty$, we find
\begin{equation}
\lim_{s \to 0}s \zeta_{B}^{+}(s)={1\over 6}\sum_{a=3}^{12}
a(a-1)(a-2)K_{a}^{(3)}=0,
\label{(7.5)}
\end{equation}
and 
\begin{equation}
\zeta_{B}^{+}(0)={5\over 4}+{1079\over 240}-{1\over 2}
\sum_{a=2}^{12}\omega(a)K_{a}^{(3)} 
+ \sum_{j=1}^{\infty}f(j)g(j)={296 \over 45},
\label{(7.6)}
\end{equation}
where
\begin{eqnarray}
\omega(a) & \equiv & {1\over 6}{\Gamma(a+1)\over \Gamma(a-2)}
\biggr[-\log(2)-{(6a^{2}-9a+1)\over 4}{\Gamma(a-2)\over \Gamma(a+1)}
\nonumber \\
&+& 2\psi(a+1)-\psi(a-2)-\psi(4)\biggr],
\label{(7.7)}
\end{eqnarray}
\begin{equation}
f(j) \equiv {(-1)^{j}\over j!}\Bigr[-1-2^{2-j}+\zeta_{R}(j-2)
(1-\delta_{j,3})+\gamma \delta_{j,3}\Bigr].
\label{(7.8)}
\end{equation}
The spectral cancellation (7.4) achieves three goals: (i) Vanishing of
$\log 2$ coefficient in Eq. (7.6); (ii) Vanishing of 
$\sum_{j=1}^{\infty}f(j)g(j)$ in Eq. (7.6); (iii) Regularity at the
origin of $\zeta_{B}^{+}$.

To cross-check our analysis, we evaluate $r_{j,+}(T)-r_{j,-}(T)$
and hence obtain $R_{j,+}(T)-R_{j,-}(T)$ for all $j$. Only $j=3$
contributes to $\zeta_{B}^{\pm}(0)$, and we find
\begin{eqnarray}
\zeta_{B}^{+}(0)&=& \zeta_{B}^{-}(0)-{1\over 24}\sum_{l=1}^{4}
{\Gamma(l+1)\over \Gamma(l-2)}\left[\psi(l+2)-{1\over (l+1)}\right]
\kappa_{2l+1}^{(3)} \nonumber \\
&=& {206 \over 45}+2={296\over 45},
\label{(7.9)}
\end{eqnarray}
in agreement with Eq. (7.6), where $\kappa_{2l+1}^{(3)}$ are the four
coefficients on the right-hand side of
\begin{equation}
{\partial \over \partial T}(R_{3,+}-R_{3,-})
=(1-T^{2})^{-4}\Bigr(80 T^{3}-24 T^{5}+32 T^{7}
-8 T^{9}\Bigr).
\label{(7.10)}
\end{equation}
Within this framework, the spectral cancellation reads as
\begin{equation}
\sum_{l=1}^{4}{\Gamma(l+1)\over \Gamma(l-2)}
\kappa_{2l+1}^{(3)}=0,
\label{(7.11)}
\end{equation}
which is a particular case of
\begin{equation}
\sum_{a=a_{\rm min}(j)}^{a=a_{\rm max}(j)}
{\Gamma((a+1)/2)\over \Gamma((a+1)/2 -j)}\kappa_{a}^{(j)}=0.
\label{(7.12)}
\end{equation}
Interestingly, the full $\zeta(0)$ value for pure gravity (i.e. including
the contribution of tensor, vector, scalar and ghost modes) is then
found to be positive \cite{F30,F31}:
\begin{equation}
\zeta(0)={142\over 45},
\label{(7.13)}
\end{equation} 
which suggests in light of (2.5) 
a quantum avoidance of the cosmological singularity driven
by full diffeomorphism invariance of the boundary-value problem for
one-loop quantum theory.

\section{Open problems}
\setcounter{equation}{0}

The DeWitt boundary condition lies at
the very heart of deep issues in quantum gravity. 
As far as we can see, the main open problems are
as follows.
\vskip 0.3cm
\noindent
(1) Among the three schemes studied in our sections 2-7, the
latter, i.e the choice of completely diff-invariant boundary
conditions on all perturbative modes, might seem the most 
satisfactory, but unfortunately the strong ellipticity of
the boundary-value problem is not fulfilled in such a case
\cite{F29,F32,F33,F34,P1,P2}. However, our analysis shows that,
{\it in the particular case} of flat Euclidean 4-space bounded
by a 3-sphere boundary, peculiar cancellations occur and the 
resulting $\zeta(0)$ value can be defined and is positive.
The deeper underlying reason might be that, in order to define
a spectral $\zeta$-function, it is sufficient to find a sector
of the complex plane free of eigenvalues of the leading symbol
of the elliptic operator under consideration (we are grateful to
Professor Gerd Grubb for correspondence about this property a
long time ago). An alternative approach might consist in 
considering non-local boundary conditions in Euclidean quantum
gravity \cite{F35,F36,F37}, or the normalizability criterion 
for the wave function of the universe \cite{F38}.
\vskip 0.3cm
\noindent
(2) The outstanding work in Ref. \cite{F09} looked for solutions of the 
quantum constraint equations in order to check validity of DeWitt's
proposal. However, although one can obtain under suitable assumptions
a formal proof of the equivalence of canonical and functional-integral
approaches \cite{F39}, DeWitt himself provided an enlightening example
of a sum over histories which does not solve the Wheeler-DeWitt
equation \cite{F40}. This remark might therefore account for the
inequivalence between our conclusions and the results in Ref. \cite{F09}.

The fascinating question of wheher our universe can be non-singular
in a semiclassical theory of quantum gravity \cite{F41} is therefore still 
waiting for a fully satisfactory answer.
\vskip 1cm

\renewcommand{\theequation}{A.\arabic{equation}}

\leftline {\bf Appendix A: the one-loop approximation}
\setcounter{equation}{0}
\vskip 0.3cm
\noindent
We are here interested in the approach to quantum field theory
in terms of Feynman functional integrals. Hence we study the amplitudes
of going from data on a spacelike surface $\Sigma_{1}$ to data
on a spacelike surface $\Sigma_{2}$. For example,
in the case of real scalar fields $\phi$ in a curved background
$M$, the data are the induced three-metric $h$ and a linear
combination of $\phi$ and its normal derivative:
$a \phi + b {\partial \phi \over \partial n}$. The latter
reduces to homogeneous Dirichlet conditions if $b=0$, and
Neumann conditions if $a=0$. Otherwise, it is a Robin boundary
condition. The quantum amplitudes are functionals of these
boundary data. On making a Wick rotation and 
using the background-field method, we may
expand both the four-metric $g$ and the field $\phi$ about
solutions of the classical field equations as 
$g=g_{0}+{\overline g}$ and $\phi=\phi_{0}+{\overline \phi}$.
However, the more general possibility remains to consider
background fields which are not solutions of any field
equations, or which are (approximate) solutions in the
asymptotic regions. The logarithm of the one-loop
functional integral $Z$ for a scalar field (in the main
body of our paper we study pure gravity, but here we focus
on scalar fields for simplicity) has an asymptotic expansion
\begin{equation}
\log(Z) \sim \log \int \mu [{\overline \phi}]
{\rm e}^{-I_{2}[{\overline \phi}]/{\hbar}}
+{\rm O}({\hbar}^{2}),
\label{(A.1)} 
\end{equation}
where $\mu$ is a suitable measure on the space
of scalar-field perturbations. The part
$I_{2}[{\overline \phi}]$ of the action which is quadratic in 
scalar-field perturbations involves a second-order elliptic
operator $\cal B$. Assuming completeness of the set
$\{ \varphi_{n} \}$ of eigenfunctions of $\cal B$, with
eigenvalues $\lambda_{n}$, the corresponding contribution to
one-loop quantum amplitudes involves an infinite product of
Gaussian integrals, i.e.
\begin{equation}
\prod_{n=n_{0}}^{\infty}
\int \mu \; dy_{n} \; {\rm e}^{-{\lambda_{n}\over 2} \; y_{n}^{2}}
={1\over \sqrt{{\rm det} \biggr({1\over 2}\pi^{-1}\mu^{-2}
{\cal B} \biggr)}}.
\label{(A.2)}
\end{equation}
In order to make sense of this infinite product of eigenvalues, one 
can use $\zeta$-function regularization. This is a
rigorous mathematical tool which relies on the spectral theorem,
according to which for any elliptic, self-adjoint, positive-definite
operator $B$, its complex powers $B^{-s}$ can be
defined. Hence its spectral $\zeta$-function is defined as in Eq. (2.4),
and the analytic continuation of the $\zeta$-function to the whole
complex-$s$ plane takes the form 
\begin{equation}
\zeta_{B}(s)=\sum_{k=-n}^{N}
{a_{k}\over \left(s+{k \over m}\right)}+\phi_{N}(s), 
\; k \not =0.
\label{(A.3)}
\end{equation}

Thus, on using analytic continuations, $\zeta_{B}(0)$ is
actually finite, and its value gives information about 
scaling properties of quantum amplitudes.
We can now be more precise and describe in detail some key
properties. The relation
\begin{equation}
{\rm det}B={\rm e}^{-\zeta'(0)}
\label{(A.4)}
\end{equation}
becomes a possible way to {\it define} the determinant of the
elliptic operator $B$ upon analytic continuation of
$\zeta_{B}(s)$. If $B$ is a second-order operator, its 
eigenvalues $\lambda_{n}$ have dimension $({\rm length})^{-2}$.
Under conformal rescaling of the metric according to
${\widehat g}=k^{2}g$, one has ${\widehat \lambda}_{n}=
\lambda_{n}/k^{2}$, and the new spectral $\zeta$-function is
${\widehat \zeta}(s)=k^{2s}\zeta(s)$. This leads to
\begin{equation}
\log \; {\rm det} {\widehat B}
=\log \; {\rm det} B - \log (k^{2}) \zeta(0),
\label{(A.5)}
\end{equation}
and hence the partition function scales as 
\begin{equation}
\log({\widehat Z})= \log(Z) +{1\over 2}\log(k^{2})
\zeta(0)+ \log ({\widehat \mu}/\mu)\zeta(0).
\label{(A.6)}
\end{equation}
The parameter $\mu$ is the one occurring in the one-loop 
semiclassical evaluation of the functional integral.
This formula allows for the more general case when the
normalization parameter $\mu$ changes under scale transformations.
One can avoid this complication by assuming that the measure 
in the functional integral is defined on scalar densities of weight
${1\over 2}$.

Equation (A.5) can also be used to deduce that the one-loop
effective action (for the scalar field) reads as
\begin{equation}
\Gamma^{(1)}={1\over 2}\log \; {\rm det} \;
{\widehat B}=-{1\over 2}\zeta'(0)-{1\over 2}\zeta(0)
\log(k^{2}).
\label{(A.7)}
\end{equation}
Note that the resulting one-loop $\langle {\rm out} \mid
{\rm in} \rangle$ amplitude is measure-dependent 
unless $\zeta(0)=0$. This is why $\zeta(0)$ is frequently
called the {\it anomalous scaling} factor.

\end{document}